\documentclass[aps,prb,reprint,longbibliography,superscriptaddress]{revtex4-2}

\usepackage{graphicx}

\usepackage[colorlinks=true,citecolor=blue,linkcolor=blue,urlcolor=blue,bookmarks=false]{hyperref}
\usepackage{CJK}
\usepackage{chemformula}
\usepackage[T1]{fontenc}
\usepackage{times}
\usepackage{subfigure}
\usepackage{amsmath,amssymb}
\usepackage{graphicx}
\usepackage{lipsum}
\usepackage{bm,dsfont}

\begin{document}
\begin{CJK}{UTF8}{bsmi}

\title{Dimensional and doping stability of Peierls charge density waves in arrays of coupled one-dimensional chains}

\author{Aitor Garcia-Ruiz (艾飛宇)}
\email{aitor.garcia-ruiz@phys.ncku.edu.tw}
\affiliation{Department of Physics and Center for Quantum Frontiers of Research and Technology (QFort), National Cheng Kung University, Tainan 70101, Taiwan}

\author{Che-Pin Hsu (許哲彬)}

\affiliation{Department of Physics and Center for Quantum Frontiers of Research and Technology (QFort), National Cheng Kung University, Tainan 70101, Taiwan}

\author{Ming-Hao Liu (劉明豪)}

\affiliation{Department of Physics and Center for Quantum Frontiers of Research and Technology (QFort), National Cheng Kung University, Tainan 70101, Taiwan}

\author{Marcin Mucha-Kruczynski}
\email{mlmk20@bath.ac.uk}

\affiliation{Department of Physics, University of Bath, Claverton Down, Bath, BA2 7AY, United Kingdom}

\begin{abstract}
The Peierls instability, the spontaneous dimerization of a one-dimensional metallic chain at half filling, is a paradigmatic mechanism for charge-density-wave (CDW) formation. Here we test its robustness under finite doping and interchain hybridization in finite-thickness arrays of identical chains. We find that the stacking geometry plays a decisive role in stabilizing CDW order away from half filling. In particular, parallel-coupled chains exhibit a bistable regime where the normal and dimerized states coexist as local minima of the total energy, while skew-coupled chains display reentrant CDW order upon doping. Our results demonstrate that even minimal models of coupled atomic chains host rich phase diagrams controlled by doping, lattice rigidity, and interchain coupling geometry.
\end{abstract}

\maketitle

\end{CJK}

\section{Introduction}

Low-dimensional materials host an exceptional variety of correlated electronic phases driven by electron-electron interactions \cite{cao_unconventional_2018, wang_correlated_2020, pantaleon_superconductivity_2023, su_superconductivity_2023, xu_tunable_2021, paschen_quantum_2021, yanase_theory_2003,little_possibility_1964}, disorder  \cite{anderson_absence_1958, li_topological_2009, groth_theory_2009, orth_topological_2016,yang_higher-order_2021,liu_topological_2022} and electron-phonon coupling \cite{kohn_image_1959, piscanec_kohn_2004, hoesch_giant_2009, ochoa_moire-pattern_2019, birkbeck_quantum_2025}. Among the most prominent, are charge density waves (CDWs), characterised by a periodic modulation of the electronic density accompanied by a lattice distortion \cite{gruner_dynamics_1988, johannes_fermi_2008, zhu_classification_2015, huang_complex_2024, chen_strong_2020, ryu_persistent_2018, luckin_controlling_2024,nakata_robust_2021}. In layered transition-metal dichalcogenides (TMDs) \cite{lin_patterns_2020, yu_unusual_2021, qi_charge_2026} and related compounds \cite{hu_coexistence_2022, luo_possible_2022}, CDWs are ubiquitous and often appear in close proximity to superconductivity and other broken symmetry states. Many materials exhibit several competing CDW orders or different lattice distortion patterns for the same CDW periodicity that lie close in energy \cite{gye_topological_2019, guster_coexistence_2019}, leading to metastability \cite{han_exploration_2015, stojchevska_stability_2018, huber_revealing_2025}, hysteretic behaviour \cite{lv_unconventional_2022, geremew_high-frequency_2020} and bistable switching \cite{mihailovic_ultrafast_2021, patel_photocurrent_2020}. While weak-coupling pictures frequently motivate CDW formation in terms of the Peierls mechanism of Fermi surface nesting and enhanced charge susceptibility at a characteristic wave vector, the observed complexity is often attributed to material-specific ingredients such as multi-orbital physics, strong coupling to the lattice and long-range interactions \cite{johannes_fermi_2008, lin_patterns_2020}. 

Peierls pointed out that, in one dimension, a lattice distortion of a half filled chain with wave vector $2k_F$ opens a gap at the Fermi points, lowering the electronic energy and rendering the metallic state unstable \cite{peierls_quantum_1955}. Equivalently, the static charge susceptibility diverges at $2k_F$, making the Peierls instability essentially unavoidable. However, this nesting-based picture is strongly tied to the kinematics of a 1D Fermi surface and generally loses predictive power in higher dimensions in the presence of more complex Fermi surfaces so that peaks in the charge susceptibility do not necessarily determine a unique ordering wave vector or CDW pattern \cite{johannes_fermi_2008}. Nevertheless, the Peierls mechanism provides a minimal and intuitive framework for studying CDW formation and it remains an important open question how much of the observed phenomenology can already emerge from minimal microscopic models.

Here, we study the fate of the Peierls mechanism in arrays of coupled atomic chains, scenario which allows us to explore the impact of another spatial dimension. Interchain hybridization discretizes the transverse motion into a set of subbands so that even weak coupling produces multiple Fermi crossings. As a result, small changes in doping can move the chemical potential across several band crossings and the CDW instability is no longer governed by a single $2k_F$ condition of an isolated chain. Instead, it depends sensitively on the subband structure and on which crossings can be gapped by the lattice distortion. Crucially, this is controlled by the stacking geometry, which determines which degeneracies are protected by symmetry and which are lifted by hybridization. For parallel stacking of chains, we find a bistable regime in which the normal and CDW states coexist as distinct local minima of the total energy. For skew stacking, we identify reentrant CDW order upon doping. Our results demonstrate that coupled atomic chains provide a remarkably simple platform in which interchain hybridization and geometry alone generate CDW competition and metastability, offering an intuitive route towards complex phenomenology within a minimal framework.

\section{1D chain under doping: Peierls mechanism}

\begin{figure}
    \centering
    \includegraphics[width=1\linewidth]{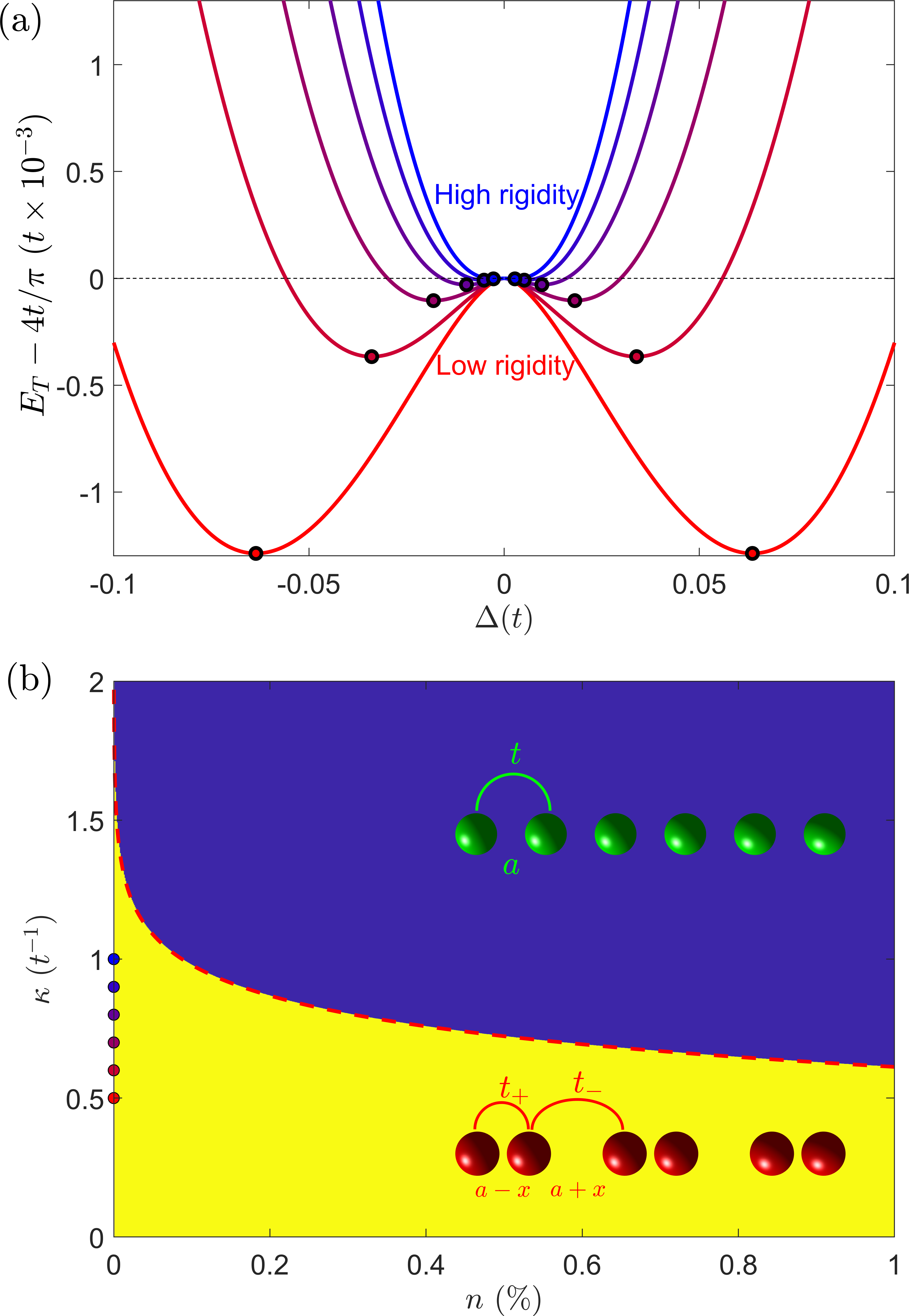}
    \caption{(a)  Total energy curves for six different values of $\kappa$ between $\kappa=0.5t^{-1}$ and $\kappa=t^{-1}$. Dots represent the position of the global minima of each curve, computed using Eq. (\ref{Eq:Delta0}). (b) Phase diagram in the doping ($n$) and rigidity ($\kappa$) parameter space. The normal (blue) and the Peierls CDW order (yellow) regions are separated by a red dashed curve given analytically in Eq. (\ref{Eq:DeltaIsMinimum}). The inset shows the two structural configurations considered in this work. The series of dots for doping $n=0$ indicates the values of rigidity used in the plots in (a), with colours of the dots matching the colours of the lines in (a).}
    \label{fig:1}
\end{figure}

We start from discussing an infinite one-dimensional chain of single-orbital atoms, distanced $a$ apart, half filled with spinless electrons. Electrons hop between nearest-neighbor sites with bare amplitude $t$. We investigate the Peierls instability towards a dimerized ground state, in which alternating bonds are shortened and elongated by a static displacement $x$, as illustrated in the insets of Fig.~\ref{fig:1}(b). The lattice is modeled by harmonic bonds, so that a distortion costs an elastic energy $Kx^2/2$ per bond, where $K$ is the spring constant. The modulation of the bond length is accompanied by a modulation of the hopping amplitude, which we parameterize as $\delta\equiv\Delta t=-\alpha x$, with $\alpha$ the electron-lattice coupling constant. The resulting electronic Hamiltonian is that of an infinite Su-Schrieffer-Heeger (SSH) chain \cite{su_solitons_1979}. Using the sublattice amplitude basis, $\{\psi^A,\psi^B\}$, where $A$ and $B$ denote the sites in the unit cell of length $2a$, the Hamiltonian reads,
\begin{align}\label{Eq:H_single}
    H_k=&
    \left(
    \begin{matrix}
        0&f_k\\
        f_k^*&0
    \end{matrix}
    \right),\quad
    f_k\equiv t_-+t_+e^{i2ak},
\end{align}
where $t_\pm=t\pm\delta/2$ and $2a$ is the unit cell length. The band energies are $\epsilon_\beta(k)=\beta\vert f_k\vert$ ($\beta=\pm1$), and determine the electronic contribution to the total energy. Equivalently, this static SSH description can be obtained from the zero-temperature electron-phonon Hamiltonian after a unitary transformation (see Appendix \ref{App:Equivalence}). Including the elastic energy, the total energy per unit cell takes the form
\begin{align}\label{Eq:E_total}
    E_{\mathrm{T}}=
    4t^2\kappa
    \Delta^2
    -\frac{4ta}{\pi}
    \int_{0}^{k_F}
    \sqrt{1-\left(1-\Delta^2\right)\sin^2(ak)}dk,
\end{align}
where we introduce the dimensionless order parameter $\Delta=\delta/2t$ and rigidity $\kappa=K/\alpha^2$ for compactness, and $k_F$ is the Fermi wavelength, which encodes the filling. Owing to electron-hole symmetry about half filling, Eq.~\eqref{Eq:E_total} applies equally to electron and hole doping.

At half filling, $k_F=\pi/2a$, and the electronic contribution reduces to the complete elliptic integral of the second kind \cite{abramowitz_handbook_1965, whittaker_course_1996}, with modulus $\sqrt{1+\Delta^2}$. Expanding Eq.~\eqref{Eq:E_total} for $\vert\Delta\vert\ll1$, we obtain
\begin{align}\label{Eq:E_total_HF}
E_{\mathrm{T}}^{\mathrm{HF}}\approx
4t^2\kappa\Delta^2-\frac{4t}{\pi}
\left\{1+\frac{1}{2}
\left[ \ln\left(
\frac{4}{\Delta}
\right)-\frac{1}{2}
\right]\Delta^2\right\}.
\end{align}
Minimizing this expression yields the optimal order parameter (see appendix \ref{App:global_minimum}),
\begin{align}\label{Eq:Delta0}
\Delta_0=
\pm\frac{4}{e}
e^{-2\pi t \kappa}.
\end{align}
Importantly, $\Delta_0$ remains nonzero for any finite rigidity $\kappa$, reflecting the Peierls instability of a one-dimensional metal at half filling \cite{peierls_quantum_1955}. The corresponding minima are indicated by dots in the energy profiles of Fig.~\ref{fig:1}(a). 

Away from half filling, Eq.~\eqref{Eq:E_total} can be expressed in terms of incomplete elliptic integrals (see Appendix \ref{App:elliptic_expansion}). Writing $k_F=\pi/2a-\epsilon/a$ ($\epsilon$ is a dimensionless measure of the doping), Eq.~\eqref{Eq:E_total} becomes
\begin{align}\label{Eq:ET}
    E_{\mathrm{T}}=&
    4t^2\kappa
    \Delta^2
    -\frac{4t}{\pi}
    \int_{0}^{\frac{\pi}{2}-\epsilon}
    \sqrt{1-\left(1-\Delta^2\right)\sin^2(\phi)}d\phi\\
    \approx&
    E_{\mathrm{T}}^{\mathrm{HF}}+
    \frac{2t}{\pi}
    \left[
    \epsilon\sqrt{\epsilon^2+\Delta^2}
    +\Delta^2
    \ln
    \left(
    \frac{\epsilon+\sqrt{\epsilon^2+\Delta^2}}{\Delta}
    \right)
    \right].\nonumber
\end{align}
For $\epsilon\neq0$, the additional logarithmic term regularizes the $\Delta\to0$ behavior of the half filled energy functional, so that $\Delta=0$ is no longer necessarily unstable. The stability of the normal state is determined by the curvature at $\Delta=0$. Requiring $\partial^2E_T/\partial\Delta^2\vert_{\Delta=0}>0$ yields the exact condition,
\begin{align}\label{Eq:DeltaIsMinimum}
    -2t\pi\kappa<
    \ln\left[\tan \left(\frac{\epsilon}{2}\right) \right]
    +
    \cos\epsilon,
\end{align}
which defines the phase boundary between the dimerized (Peierls) state and the normal state in Fig.~\ref{fig:1}(b), where the doping $n$ is expressed as the percentage of excess electron density relative to half filling. Note that throughout this work we focus on the weak-coupling Peierls regime, characterized by low doping $\epsilon\ll1$ and weak dimerization $\Delta\ll 1$. Under these conditions, the conventional dimerized CDW with wave vector $2k_F=\pi/a$ remains the dominant instability, while competing higher -periodicity or incommensurate charge modulations \cite{ohmi_commensurate-incommensurate_1977, yamamoto_commensurate-incommensurate_1978} appearing at higher dopings can be neglected.

\section{1D coupled chains}

\begin{figure}
    \centering
    \includegraphics[width=1\linewidth]{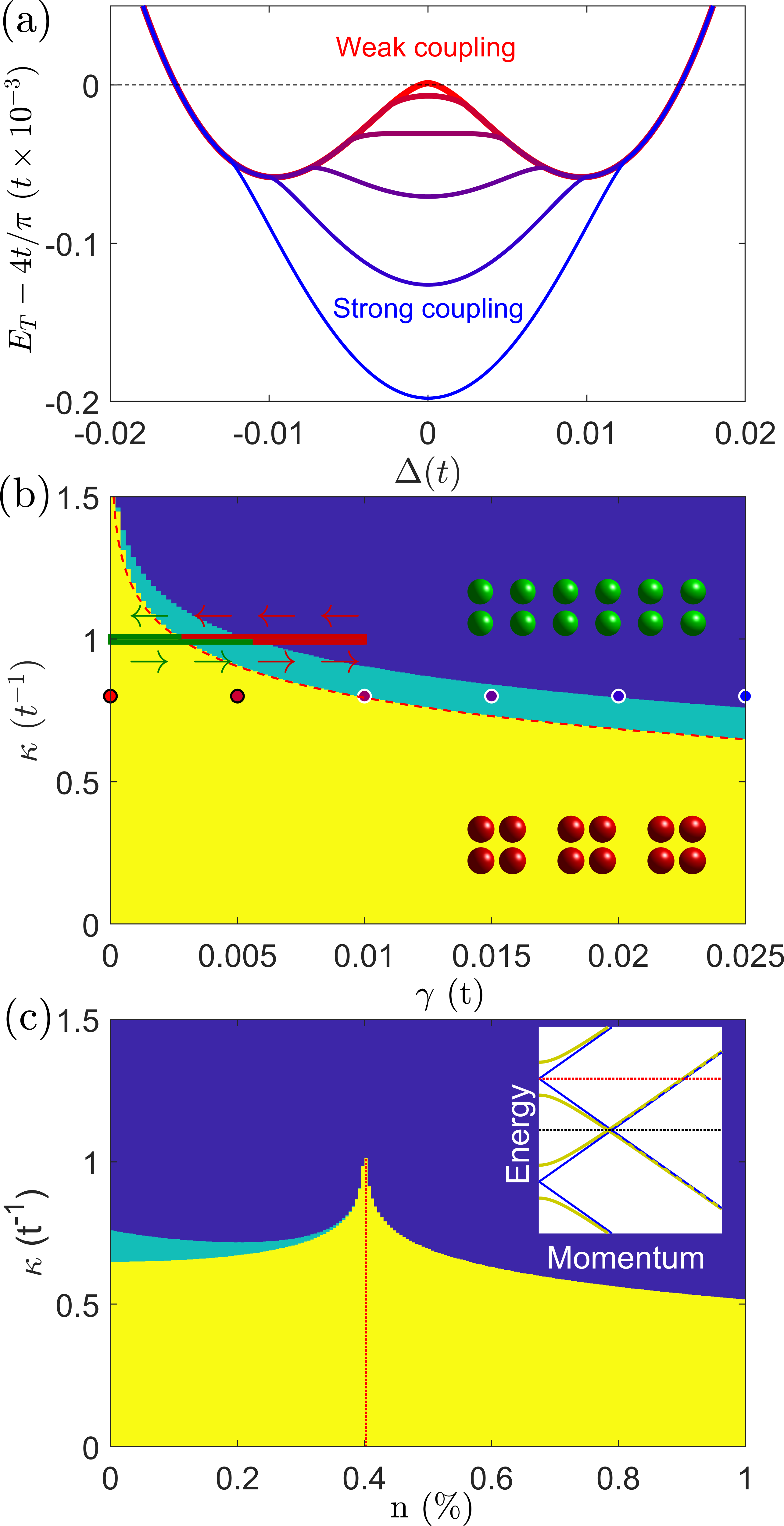}
    \caption{(a) Total energy of two parallel-coupled atomic chains for different values of $\gamma$. As the coupling increases, the total energy transitions from two to three minima before $\Delta=0$ establishes as the global minimum. (b) Phase diagram of two half-filled parallel-coupled chains. In between the phase separation, there is a green region denoting bistability. A closed path at constant $\kappa$ is marked with green and red lines and green and red arrows indicating the direction of circulation. The colour of the line and corresponding arrow reflects the state of the chains (red - normal metal, green - CDW). (c) Phase diagram as a function of doping for $\gamma=0.025t$. The level of doping at $\sim0.4\%$ that enhances the CDW order corresponds to the Fermi level at which a gap is opened in the band structure at the edge of the Brillouin zone $k=-\pi/a$, as shown in the inset. }
    \label{fig:2}
\end{figure}

\begin{figure*}
    \centering
    \includegraphics[width=1\linewidth]{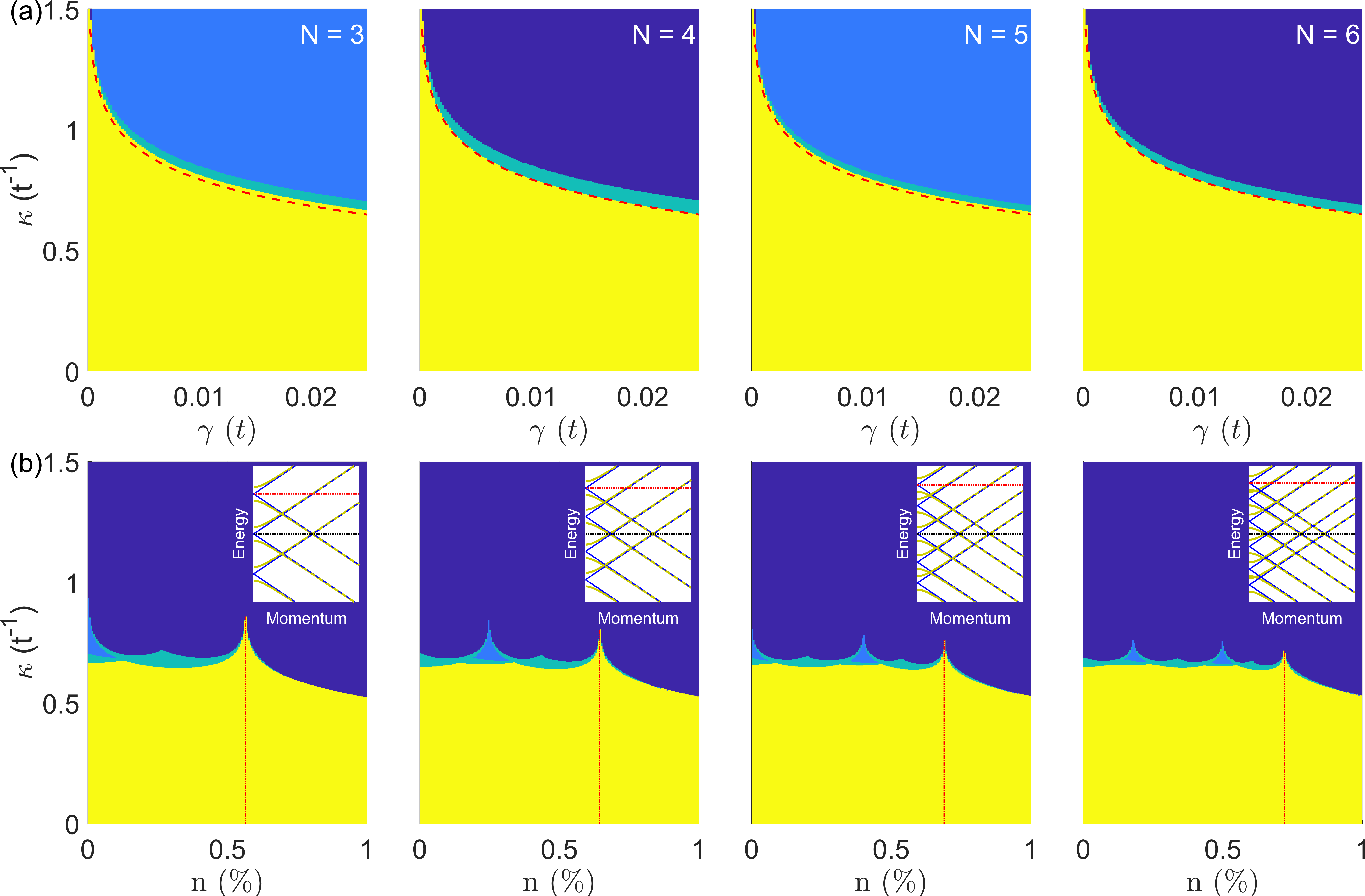}
    \caption{(a) Phase diagram of half-filled parallel-coupled chains as a function of interchain coupling $\gamma$ and rigidity $\kappa$, with number of chains ranging from $N=3$ (leftmost) to $N=6$ (right most). Light blue represents a weak CDW order due to gapping of zero-energy bands, that must exist for odd $N$, electron-hole symmetric spectra. All even-number chains share the same CDW/bistable regime phase boundary (see appendix \ref{App:2N}). (b) Phase diagram as a function of doping and rigidity, with $\gamma=0.025t$. All subbands open a gap at the edge of the Brillouin zone $k=-\pi/2a$. The doping level at these energy points maximizes the energy reduction of the system and enhances the CDW order.}
    \label{fig:3}
\end{figure*}
For an array of $N$ identical, weakly-coupled chains, the Hamiltonian, in the basis of orbital amplitude $\Psi^{\dagger}= \{  \psi_1^{A\,\dagger}, \psi_1^{B\,\dagger}, \cdots, \psi_N^{A\,\dagger}, \psi_N^{B\,\dagger}\}$, takes the block-tridiagonal form
\begin{align}\label{Eq:H_N}
    \mathcal{H}=
    \left(
    \begin{matrix}
        H_k&T_{1,2}&0&\cdots&0\\
        T_{1,2}^{\dagger}&H_k&T_{2,3}&\cdots&0\\
        0&T_{2,3}^\dagger&H_k&\cdots&0\\
        \vdots&\vdots&\vdots&\ddots&\vdots\\
        0&0&0&\cdots&H_k
    \end{matrix}
    \right).
\end{align}
Above, the diagonal $2\times2$ blocks $H_k$ describe a single chain and the off-diagonal blocks $T_{n,n+1}$ encode the microscopic coupling between neighbouring chains. In the following, we consider two types of interchain coupling: parallel and skew.  {In this work, we focus on intra-chain dimerization as the dominant CDW channel. The effects of additional transverse modulations of the interchain hopping amplitudes are discussed in Appendix \ref{App:distortions}, where we show that they do not qualitatively modify the phase diagram in the weak-coupling regime considered here.

\subsection{Parallel coupling}
For parallel coupling, the chains are placed in an array such that each consecutive chain is a copy of the previous chain translated perpendicularly to the chain direction by the same interchain distance. The interchain coupling matrix is $T_{n,n+1}=\gamma\mathbb{I}_2$ in Eq. (\ref{Eq:H_N}), where $\gamma$ is the interchain hopping amplitude. Since the interchain hopping is proportional to identity in sublattice space, the Hamiltonian can be written as 
\begin{align}
    \mathcal{H}= \mathbb{I}_N\otimes H_k+H_\perp\otimes\mathbb{I}_2,
    \nonumber
\end{align}
where $H_\perp$ is the $N\times N$ tridiagonal matrix describing the nearest neighbor hopping between chains. Diagonalizing $H_\perp$ yields $N$ decoupled Hamiltonians,
\begin{align}
\mathcal{H}\to\mathbb{I}_N\otimes H_k + \mathrm{diag}(\mu_1,\cdots,\mu_N) \otimes \mathbb{I}_2, \\ \mu_n=2\gamma\cos\left(\frac{n\pi}{N+1}\right), \quad n=1,\cdots,N.\nonumber
\end{align}
Accordingly, the spectrum consists of rigidly shifted replicas of the single-chain dispersion, $\beta\vert f_k\vert+\mu_n$, that is, the parallel-coupled system is energetically equivalent to $N$ independent 1D chains, each subject to an effective shift of the chemical potential, $\mu_n$.

The total energy of two parallel-coupled chains as a function of the dimerization amplitude $\Delta$, for different values of the coupling strength $\gamma$, is shown in Fig. \ref{fig:2}(a). The corresponding band structure consists of two 1D atomic chain dispersions, shifted by $\mu=\pm \gamma$. For sufficiently strong interchain coupling, the undistorted configuration ($\Delta=0$) becomes the global minimum once the condition in Eq.~(\ref{Eq:DeltaIsMinimum}) is met, while for weak interchain couplings the system features a double-well energy profile. For intermediate values of the interchain coupling, a local minimum at $\Delta=0$ emerges and coexists with another one at finite values of $\Delta=\pm\Delta_0$. In this situation, the system is bistable, and the complete phase diagram is presented in Fig. \ref{fig:2}(b), where we show the region of bistability in green.  {The existence of this regime is independent of the specific value of rigidity, and it should give rise to hysteretic behavior when the inter-chain coupling $\gamma$ is cycled across the bistable region, as indicated by the arrows in the inset. This reflects the coexistence of competing local minima in the total energy landscape. Experimentally, hydrostatic pressure and uniaxial strain in quasi-one-dimensional CDW materials have been demonstrated to tune the transverse orbital overlap between chains \cite{konczewicz_pressure_1982, lear_stress_1984,kuh_NbSe3_1998}, thereby effectively modifying the inter-chain hybridization parameter $\gamma$ in our model and providing a potential route to probe the bistable phase.

The stability of the CDW state under doping is also qualitatively modified compared to a single chain, where doping suppresses the Peierls distortion. Figure ~\ref{fig:2}(c) shows the phase diagram as a function of doping and rigidity. Initially, departure from half filling strengthens the CDW order up to a characteristic scale set by the interchain coupling, $E_F=\vert\gamma\vert$, where the Fermi level lies close to the band edge and the opening of a Peierls gap maximally lowers the electronic energy. At larger doping, the bistable regime disappears, and the system recovers the single-chain trend where doping weakens the CDW order.

Figure~\ref{fig:3}(a) generalizes the half filling phase diagram to stacks of $N=3,\cdots,6$ parallel-coupled chains. Bistability persists for all $N$. A key even-odd effect is also present: stacks with odd $N$ never reach a fully metallic normal state. This follows from the transverse spectrum $\mu_n$, which for odd $N$ must contain a mode with $\mu_{(N+1)/2}=0$. One pair of effective single-chain bands, therefore, remains exactly at half filling, and retains a Peierls instability, even when the other modes are shifted away. This produces the weaker CDW phase with a marginal lattice distortion, shown in light blue. By contrast, even-$N$ stacks exhibit phase diagrams closely analogous to the $N=2$ case (see appendix \ref{App:2N} for further discussion).

The corresponding doping-dependent phase diagrams for $\gamma=0.025t$ are shown in Fig.~\ref{fig:3}(b). Once the upper bands become significantly populated, all stacks display a monotonic suppression of CDW order with increasing doping. At lower doping, however, the phase boundaries develop a sequence of lobes that reflects the discrete set of transverse band edges. This behavior is consistent with the band structure shown in the insets: when the chemical potential lies close to a band degeneracy at the zone edge (red dotted line in the insets), the Peierls distortion is enhanced, leading to local strengthening of the CDW order.

\subsection{Skew coupling}

\begin{figure}
    \centering
    \includegraphics[width=1\linewidth]{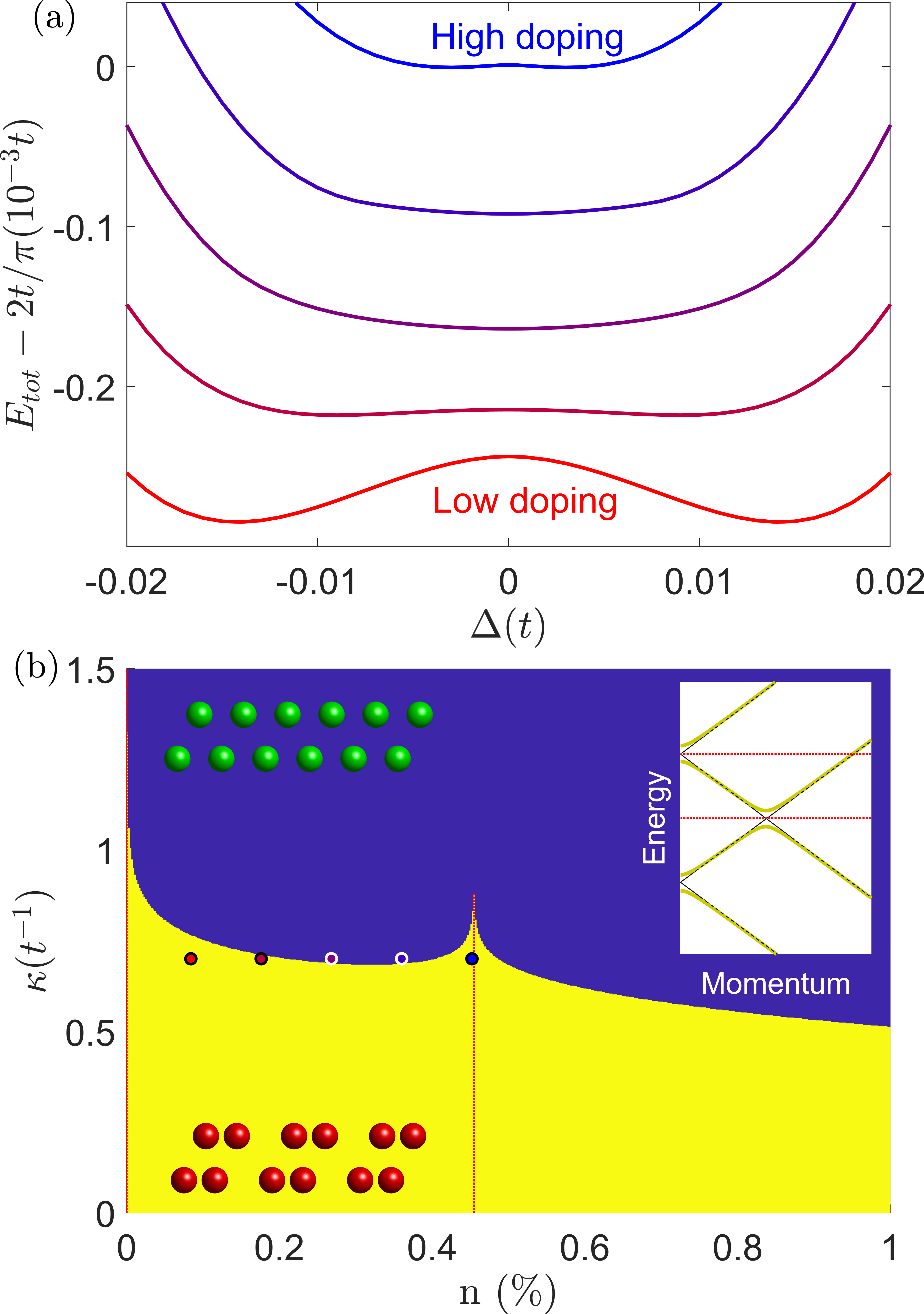}
    \caption{(a) Total energy $E_{\mathrm{tot}}(\Delta)$ of two skew-coupled chains for $\kappa=0.7t^{-1}$ and $\gamma=0.02t$, shown for several representative electron dopings. The evolution of the global minimum, from $\Delta\neq0$ to $\Delta=0$ and back to $\Delta\neq0$, demonstrates reentrant CDW order. (b) Phase diagram as a function of electron doping $n$ and rigidity $\kappa$. Inset: band structure near the Brillouin-zone edge, illustrating that CDW order is enhanced when the chemical potential lies close to a spectral gap (in particular at $n=0$ and $n\simeq0.45\%$).}
    \label{fig:4}
\end{figure}
\begin{figure*}
    \centering
    \includegraphics[width=1\linewidth]{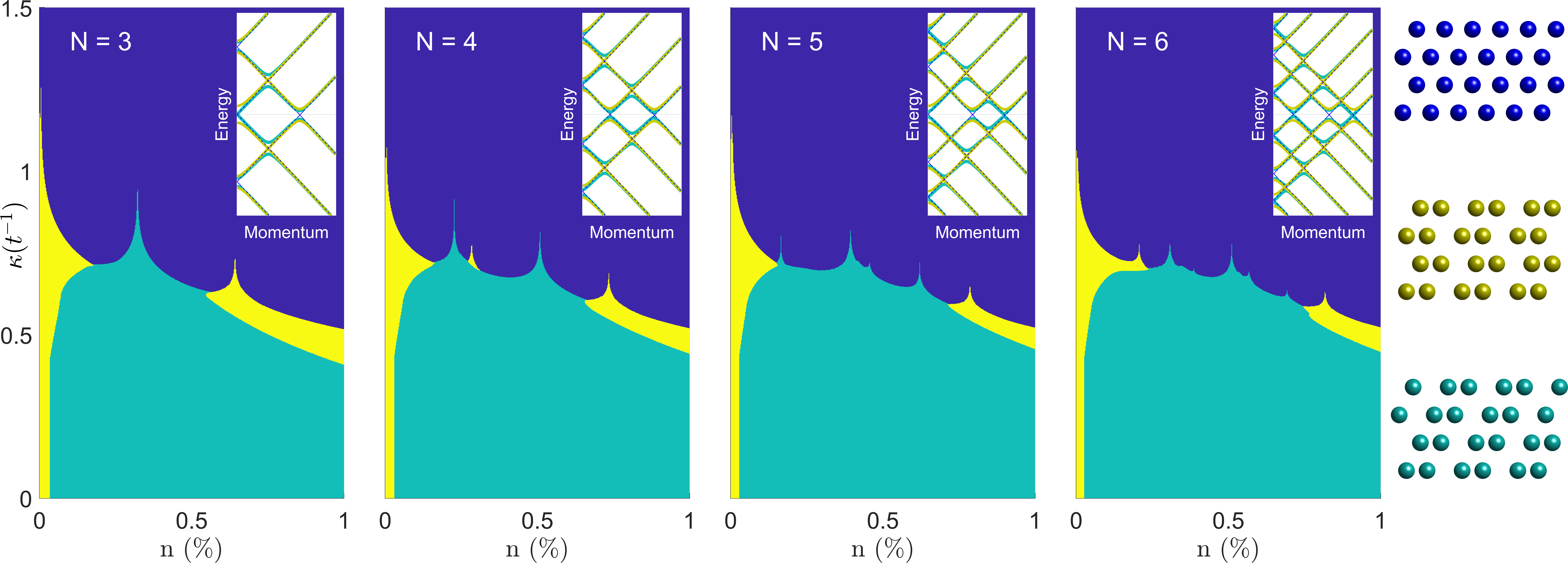} 
    \caption{ Phase diagrams of skew-coupled stacks with $N=3,\ldots,6$ chains for $\gamma=0.02t$, shown as a function of electron doping $n$ and rigidity $\kappa$.
The colors denote the ground state obtained from the global minimum of $E_{\rm tot}(\Delta)$: normal state ($\Delta=0$), zig-zag CDW, and nematic CDW. Insets: corresponding band structures near the Brillouin-zone edge $k=-\pi/2a$. At low doping, the zig-zag configuration is favored, while at intermediate doping the nematic configuration becomes energetically preferred over an extended parameter range, consistent with the relative size and location of the spectral gaps.}
    \label{fig:5}
\end{figure*}

We now consider a skew-coupled geometry in which sites of chain $n+1$ lie at the midpoints between consecutive sites of chain $n$. Each atom, therefore, hybridizes with its two nearest neighbors in the adjacent chain with equal amplitude $\gamma$. In contrast to parallel coupling, for $N\geq3$, this geometry admits two distinct commensurate CDW stacking configurations, depending on the relative phase of the dimerization between neighboring chains. We refer to the configuration in which next-nearest chains are in phase as zig-zag (zzg) and to the configuration in which they are out of phase as nematic (nem).

In Eq.~(\ref{Eq:H_N}), the corresponding interchain hopping blocks take the form
\begin{subequations}
\begin{align}
    T_{n,n+1}^{\mathrm{zzg}}=&
    \left\{
    \begin{matrix}        
    \gamma\left(
    \begin{matrix}
        1&e^{2iak}\\
        1&1
    \end{matrix}
    \right),\quad \mathrm{for}  \,\, n \,\, \mathrm{odd}
    \\
    \gamma\left(
    \begin{matrix}
        1&1\\
        e^{-2iak}&1
    \end{matrix}
    \right),\quad \mathrm{for}  \,\, n \,\, \mathrm{even}
    \end{matrix}
    \right.\\    
    T_{n,n+1}^{\mathrm{nem}}=&\,
    \gamma\left(
    \begin{matrix}
        1&e^{2iak}\\
        1&1
    \end{matrix}
    \right),\quad \forall n
\end{align}
\end{subequations}
Unlike the parallel-coupled case, this Hamiltonian cannot be reduced to a direct sum of independent shifted single-atomic chains. We therefore determine the total energy and the equilibrium dimerization numerically.

Skew coupling qualitatively modifies the band structure by lifting the sublattice-exchange symmetry of the parallel-stacked geometry and producing additional band crossings near the Fermi level. While the spectrum of the undistorted phase remains gapless at these crossings, a finite dimerization $\Delta\neq0$ breaks sublattice symmetry and opens gaps at all of them. As a result, the competition between the elastic cost and the electronic energy gain from dimerization depends sensitively on the position of the chemical potential. Therefore, away from half filling, CDW order is not necessarily monotonic in doping. Instead, the CDW can be enhanced when the Fermi level approaches a band edge, and can even become reentrant if multiple gaps occur at different energies. This behaviour is illustrated in Fig.~\ref{fig:4}(a), which shows the total energy $E(\Delta)$ for a skew-coupled double chain at several representative doping levels. At low doping (red curve), the energy is minimized at a finite distortion, while intermediate dopings suppress CDW. Upon further doping (blue line), the minimum reappears at $\Delta\neq0$, demonstrating reentrant CDW order. The corresponding phase diagram is shown in Fig.~\ref{fig:4}(b). The reentrance occurs when the chemical potential approaches a higher-energy spectral gap near the Brillouin-zone edge (inset), which restores the electronic energy gain from dimerization around $n\simeq0.45\%$.

Finally, Fig.~\ref{fig:5} shows the phase diagrams for skew-coupled stacks with $N=3,\cdots,6$ chains. In these systems the existence of two distinct CDW stackings produces a richer phase structure than in the parallel-coupled case. Nevertheless, the phase boundaries can be consistently interpreted in terms of the location and magnitude of band gaps in the electronic dispersion. At low doping, the zig-zag configuration is always preferred. At half filling, the relevant gap occurs at the Brillouin-zone edge and, in the zig-zag configuration, opens linearly with $\Delta$. In contrast, in the nematic configuration this gap is symmetry-suppressed at linear order and is generated only at second order, scaling as $\Delta^2$ (see Appendix \ref{App:zzg_vs_nem}). At moderate doping, however, the nematic stacking can open larger gaps at the relevant Fermi level, leading to an extended nematic-CDW region. At higher doping, the zig-zag phase reappears once the chemical potential approaches the next set of band edges. This sequence of transitions is consistent with the multiple avoided crossings visible in the band-structure insets of Fig.~\ref{fig:5}.

\section{Conclusions}
In summary, we have studied the robustness of the Peierls instability under finite doping and interchain hybridization in finite-thickness arrays of identical atomic chains. We find that the interplay between doping, lattice rigidity, and transverse coupling produces a surprisingly rich phenomenology that goes well beyond the expectation of monotonic CDW suppression with doping. In parallel-coupled stacks, we identify a bistable regime where the normal and CDW-ordered configurations coexist as distinct local minima of the total energy. In skew-coupled geometries, multiple avoided crossings near the Fermi level lead to nonmonotonic doping dependence and signatures of reentrant CDW order.

This behavior is reminiscent of the complex CDW phase competition observed in quasi-two-dimensional materials, where multiple commensurate and incommensurate modulations may lie close in energy and evolve non-trivially with external control parameters \cite{bin_subhan_charge_2021, leininger_competing_2011, tymoshenko_charge-density-wave_2025}. 

More broadly, our results establish coupled atomic chains as a minimal platform that naturally bridges between the analytically transparent Peierls mechanism in one dimension and the richer energy landscapes typical of layered CDW compounds.  {In this context, stacking geometry is an intrinsic structural property of the system that determines the resulting electronic landscape, while the interchain coupling could used as an effective tuning parameter, for example via hydrostatic pressure \cite{padhi_pressure-induced_2019,zocco_pressure_2015} or uniaxial strain \cite{kundu_charge_2024,zhang_tunable_2025}, thereby enabling external control over the competition between normal and CDW-ordered states. Our analysis is carried out within a controlled regime of weak-to-moderate dimerization and low doping, where the SSH-type description remians valid, and focuses on the competition between electron-lattice coupling, interchain hybridization and stacking geometry, in isolation from additional correlation effects. We list in Table I examples of materials to which such an approach should be applicable. In other systems, the impact of additional effects like the electron-electron interaction might be important to the extent that the phase diagrams are reshaped [58] or new phases (for example, Luttinger liquid) appear [59]. Incorporating electron-electron interaction within an extended Hubbard model [60] would therefore be a natural extension of our work to treat it on an equal footing with the Peierls instability.

In realistic materials, the phases described here can compete with electron-electron interactions, which can further reshape the phase diagram \cite{kang_band-selective_2021}  and, in some cases, drive Luttinger-liquid dominated regimes \cite{zhu_imaging_2022}. Incorporating electron-electron interactions within an extended Hubbard model \cite{nakamura_mechanism_1999} would therefore be a natural extension of our work to treat this competition on equal footing.

\begin{table}[t]
\caption{
Representative energy scales reported for quasi-one-dimensional materials estimated from ARPES measurements or first-principles calculations. The intrachain hopping $t$, interchain coupling $\gamma$ are inferred from the longitudinal and transverse electronic bandwidths, respectively, while $\delta$ is obtained from the reported CDW gap energy. The resulting dimensionless ratios indicate that these materials predominantly occupy the regime $\gamma/t\ll1$ explored in our model.}

\label{tab:materials}
\begin{ruledtabular}
\begin{tabular}{lccccc}
System 
& $t$ (eV)
& $\gamma$ (meV)
& $\delta$ (meV)
& $\gamma/t$
& $\delta/t$
\\
\hline
 K$_{0.3}$MoO$_3$ \cite{kang_band-selective_2021}
& $1$
& $50$
& $42$
& $\sim10^{-2}$
& $\sim10^{-2}$
\\

NbSe$_3$ \cite{nicholson_dimensional_2017}
& $0.7$
& $27$
& $50$
& $\sim10^{-2}$
& $\sim10^{-2}$
\\

CuTe \cite{nhat_quyen_three-dimensional_2024}
& $1$
& $10$
& $100$
& $\sim10^{-2}$
& $\sim10^{-1}$
\\

(TaSe$_4$)$_2$I \cite{lin_unconventional_2024}
& $1$
& $40$
& $200$
& $\sim10^{-2}$
& $\sim10^{-1}$



\\

In atoms on Si(111) \cite{cheon_chiral_2015}
& $0.4$
& $80$ 
& $100$
& $\sim10^{-1}$
& $\sim10^{-1}$
\\

\end{tabular}
\end{ruledtabular}
\end{table}

\section*{Data availability}

The MATLAB codes and data used to generate the results presented in this work are publicly available at
\url{https://github.com/Garcia-Ruiz-Physics/CDW_in_coupled_chains/}.

\begin{acknowledgements}
We thank Tse-Ming Chen, Ming-Wen Chu and Hsin-Zon Tsai for useful discussions. A.\ G.-R., C.-P.\ H., and M.-H.\ L.\ gratefully acknowledge National Science and Technology Council (NSTC) of Taiwan (Grant No.\ 112-2112-M-006-019-MY3, No.\ 113-2918-I-006-004, No.\ 114-2811-M-006-038, and No.\ 114-2112-M-006-029-MY3) for financial support. 
\end{acknowledgements}
\newpage

\appendix
\section{Connection to mean-field theory}
\label{App:Equivalence}
Here, we demonstrate that the Hamiltonian employed in Eq. (\ref{Eq:H_single}) is equivalent to that derived from the mean-field theory. The general form of the Hamiltonian of a one-dimensional chain of single-orbital atoms with inter-atomic spacing $a$, occupied by spinless electrons, in the framework of second quantization reads
\begin{align}
    H_{1D}=&
    H_{\mathrm{el}}+
    H_{\mathrm{ph}}+
    H_{\mathrm{el-ph}}\\
    H_{\mathrm{el}}=&
    \sum_{k}\epsilon_k
    \hat{c}^{\dagger}_{k}
    \hat{c}_{k}\nonumber\\
    H_{\mathrm{ph}}=&
    \sum_{q}\hbar \omega_q
    \hat{b}^{\dagger}_{-q}
    \hat{b}_{q}\nonumber\\
    H_{\mathrm{el-ph}}=&
    \sum_{\langle k, q\rangle}
    g(k,q)
    \hat{c}^{\dagger}_{k+q}
    \hat{c}_k
    \left(
    \hat{b}^{\dagger}_{-q}+
    \hat{b}_{q}
    \right)\nonumber,
\end{align}
where $\hat{c}^{\dagger}_{k}$ ($\hat{c}_{k}$) and $\hat{b}_q^{\dagger}$ ($\hat{b}_q$) are the operators that create (destroy) an electron and a phonon with momentum $k$ and $q$, respectively. Above, we also introduce the electronic dispersion $\epsilon_k=2t\cos(ka)$ of the electron gas, with $t$ being the hopping constant. At half filling, the Fermi surface is nested by the vector $2k_F=\pi/a$, and at zero temperature, the Peierls instability is associated with condensation of the phonon mode at $q=\pm2k_F$. To reduce the Hamiltonian above into a bilinear form, we assume the electron-phonon coupling $g(k,q)=g$ is constant, and define the order parameter $\eta=g/\sqrt{L}\langle \hat{b}_{2k_F} + \hat{b}_{-2k_F}^\dagger \rangle$. After mean-field decoupling and diagonalization in the reduced Brillouin zone, the mean-field couples electronic states at $k$ and $k+\pi/2a$, yielding the effective two-component basis of right/left movers, $\left(\hat{c}_{+,k}^{\dagger}; \hat{c}_{-,k}^{\dagger}\right)$. The resulting Hamiltonian takes the form \cite{solyom_fundamentals_2010}
\begin{align}\label{Eq:H_MF}
    \mathcal{H}_{1D}^\mathrm{MF}=&
    \left(
    \begin{matrix}
        \epsilon_k^{+}&\eta\\
        \eta&\epsilon_k^{-}
    \end{matrix}
    \right)=
    \mathbf{\sigma}\cdot
    \mathbf{d}_\mathrm{MF},\\
    \mathbf{d}_\mathrm{MF}=&
    \left(
    \eta,
    0,
    \epsilon_k    
    \right)=\sqrt{\epsilon_k^2+\eta^2}[\sin(\varphi_k),0,\cos(\varphi_k)],\nonumber\\
    \sin(\varphi_k)=&\frac{\eta}{\sqrt{\epsilon_k^2+\eta^2}},\quad
    \cos(\varphi_k)=\frac{\epsilon_k}{\sqrt{\epsilon_k^2+\eta^2}}
    \nonumber\\
\end{align}
Above, we chose a gauge in which the order parameter is real. In the same language, we can re-express the Hamiltonian in Eq. (\ref{Eq:H_single}) as
\begin{align}
    \mathcal{H}_\mathrm{SSH}=&
    \mathbf{\sigma}\cdot\mathbf{d}_\mathrm{SSH}\\
    \mathbf{d}_\mathrm{SSH}=&\left[
    \mathrm{Re}(f_k),-\mathrm{Im}(f_k),0\nonumber
    \right],\\
    =&\vert f_k\vert \left[
    \cos(\theta_k),-\sin(\theta_k),0\nonumber
    \right],
\end{align}
where $f_k=\sqrt{\epsilon_k^2+\delta^2\sin^2(ka)}\,e^{i\theta_k}\approx\sqrt{\epsilon_k^2+\delta^2}\,e^{i\theta_k}$ in the vicinity of the Brillouin zone edge. Both $H_\mathrm{MF}$ and $H_\mathrm{SSH}$ are traceless, and the 3D vectors $\mathbf{d}_\mathrm{MF}$ and $\mathbf{d}_\mathrm{SSH}$ have the same length, thus there must exist a rotation $R(k)\in SO(3)$, such that $\mathcal{U} \left(\mathbf{\sigma}\cdot\mathbf{d}_\mathrm{SSH}\right) \mathcal{U}^\dagger=\mathbf{\sigma}\cdot\mathbf{d}_\mathrm{MF}$. To note, this unitary transformation is allowed to be $k$-dependent, reflecting the freedom to choose a $k$-dependent Bloch basis in the two-component sublattice space. We can construct this unitary transformation by composing two rotations: the first one being a rotation in the $xy$-plane that removes the y-component, 
\begin{align}
    \mathcal{U}_1H_\mathrm{SSH}\mathcal{U}_1^\dagger=\vert f_k\vert \sigma_x,\\
    \mathcal{U}_1(k)=\exp(i\frac{\theta_k}{2}\sigma_z),\nonumber
\end{align}
and the second one being a rotation of the Hamiltonian by an angle $\varphi_k$,
\begin{align}
    \mathcal{U}_2
    \vert f_k\vert\sigma_x 
    \mathcal{U}_2^\dagger=&
    \vert f_k \vert
    \left[
    \sin\varphi_k\sigma_x+
    \cos\varphi_k\sigma_z
    \right]\equiv H_\mathrm{MF},\\
    \mathcal{U}_2=&\exp(-i\frac{\varphi_k}{2}\sigma_y).\nonumber
\end{align}
Therefore, the unitary matrix $\mathcal{U}(k)=\mathcal{U}_2(k)\mathcal{U}_1(k)$ transforms the SSH Hamiltonian in Eq. (\ref{Eq:H_single}), written in the basis of A/B dimer sublattices, into the Hamiltonian in Eq. (\ref{Eq:H_MF}), written in the basis of right/left movers. The two Hamiltonians are unitarily equivalent after identifying the MF order parameter $\eta$ with the low-energy dimerization gap $\delta$.

\section{Analytical minimum of the total energy}
\label{App:global_minimum}

At half filling, the total energy per unit cell ($2a$) is given by the sum of elastic and band contributions,
\begin{align}
\frac{E_{\mathrm{tot}}}{N}
=
\frac{K\delta^2}{\alpha^2}
-\frac{1}{N}\sum_{k=-\frac{\pi}{2a}}^{\frac{\pi}{2a}}
\sqrt{t_+^2+t_-^2+2t_+t_-\cos(2ka)},
\end{align}
In the thermodynamic limit,
$\frac{1}{N}\sum_k \to \frac{2a}{2\pi}\int dk$, the band contribution becomes
\begin{align}
\frac{E_{\mathrm{band}}}{N}
&=
-\frac{a}{\pi}\int_{-\frac{\pi}{2a}}^{\frac{\pi}{2a}} dk\,
\sqrt{t_+^2+t_-^2+2t_+t_-\cos(2ka)}\\
&=
-\frac{1}{\pi}\int_{-\pi/2}^{\pi/2} d\phi\,
\sqrt{4t^2\cos^2\phi+\delta^2\sin^2\phi},
\end{align}
where we define $\phi=ka$. Introducing the dimensionless dimerization parameter $
\Delta\equiv {\delta}/{2t}$, we obtain
\begin{align}
\frac{E_{\mathrm{band}}}{N}
=&
-\frac{2t}{\pi}\int_{-\frac{\pi}{2}}^{\frac{\pi}{2}} d\phi\,
\sqrt{1-(1-\Delta^2)\sin^2\phi}\\
=&
-\frac{4t}{\pi}\,
\mathcal{E}\!\left(\sqrt{1-\Delta^2}\right),
\end{align}
where $\mathcal{E}(k)$ denotes the complete elliptic integral of the second kind. It is convenient to write the total energy in the dimensionless form
\begin{align}
\frac{E_{\mathrm{tot}}}{N}
=
4t^2\kappa\,\Delta^2
-\frac{4t}{\pi}\,
\mathcal{E}\!\left(\sqrt{1-\Delta^2}\right).
\label{Eq:Etot_elliptic}
\end{align}
For $\Delta\ll 1$, that is, close to the uniform chain, the elliptic integral admits the standard expansion
\begin{align}
\mathcal{E}\!\left(\sqrt{1-\Delta^2}\right)
=
1+\frac{\Delta^2}{2}
\left[
\ln\!\left(\frac{4}{\Delta}\right)-\frac{1}{2}
\right]
+O(\Delta^4\ln\Delta).
\label{Eq:E_expansion}
\end{align}
Substituting Eq.~(\ref{Eq:E_expansion}) into Eq.~(\ref{Eq:Etot_elliptic}) yields
\begin{align}
\frac{E_{\mathrm{tot}}}{N}
\simeq
4t^2\kappa\,\Delta^2
-\frac{4t}{\pi}
\left\{
1+\frac{\Delta^2}{2}
\left[
\ln\!\left(\frac{4}{\Delta}\right)-\frac{1}{2}
\right]
\right\}.
\label{Eq:Etot_smallDelta}
\end{align}

The stationary points satisfy $\partial_\Delta(E_{\mathrm{tot}}/N)=0$,
\begin{align}
8t^2\kappa\,\Delta
-\frac{4t}{\pi}\,
\Delta
\left[
\ln\!\left(\frac{4}{\Delta}\right)-1
\right]=0.
\end{align}
Besides the trivial solution $\Delta=0$, this equation admits a nonzero solution,
\begin{align}
\Delta_0=\frac{4}{e}\,e^{-2\pi t\kappa}.
\end{align}

\section{Expansion of the incomplete elliptic integral}
\label{App:elliptic_expansion}

Here we derive the analytical expression for the incomplete elliptic integral appearing in Eq.~(\ref{Eq:ET}). We consider
\begin{align}
\tilde{\mathcal{E}}(\epsilon,\Delta)
\equiv
\int_{0}^{\frac{\pi}{2}-\epsilon} d\phi\,
\sqrt{1-\left(1-\Delta^2\right)\sin^2\phi},
\end{align}
which can be written as a complete elliptic integral of the second kind minus a tail contribution,
\begin{align}
\tilde{\mathcal{E}}(\epsilon,\Delta)
=
\mathcal{E}\!\left(\sqrt{1-\Delta^2}\right)-I(\epsilon,\Delta),
\end{align}
where $\mathcal{E}(k)\equiv \int_{0}^{\pi/2} d\phi\,\sqrt{1-k^2\sin^2\phi}$ and
\begin{align}
I(\epsilon,\Delta)
&\equiv
\int_{\frac{\pi}{2}-\epsilon}^{\frac{\pi}{2}} d\phi\,
\sqrt{1-\left(1-\Delta^2\right)\sin^2\phi}.
\end{align}
Introducing $\phi\to \frac{\pi}{2}-\phi$ in the tail integral yields
\begin{align}
I(\epsilon,\Delta)
=
\int_{0}^{\epsilon} d\phi\,
\sqrt{1-\left(1-\Delta^2\right)\cos^2\phi}.
\end{align}

For $\epsilon\ll 1$, the integrand can be expanded to leading order as $\left[ 1-\left(1-\Delta^2\right)\cos^2\phi \right]^{1/2} \approx \left[ \Delta^2+\left(1-\Delta^2\right)\phi^2 \right]^{1/2}$. This approximation amounts to dropping the term $-(1-\Delta^2)\phi^4/3$. Requiring such term to remain small compared to $\Delta^2$ over $\phi\in[0,\epsilon]$ implies the additional condition $\epsilon\ll \Delta^{1/2}$. With this truncation, the integral reads
\begin{align}
I(\epsilon,\Delta)
&\simeq
\int_{0}^{\epsilon} d\phi\,
\sqrt{\Delta^2+\left(1-\Delta^2\right)\phi^2}
\nonumber\\
&=\sqrt{1-\Delta^2}\int_{0}^{\epsilon} d\phi\,
\sqrt{\phi^2+C^2},
\end{align}
where we defined $C^2\equiv \frac{\Delta^2}{1-\Delta^2}$. The remaining integral can be evaluated analytically using hyperbolic trigonometric functions, 
\begin{align}
I(\epsilon,\Delta)
\approx&
\frac{\sqrt{1-\Delta^2}}{2}
\left[
\epsilon\sqrt{\epsilon^2+C^2}
+C^2\ln\!\left(
\frac{\epsilon+\sqrt{\epsilon^2+C^2}}{C}
\right)
\right]\nonumber\\
\approx&
\frac{1}{2}
\left[
\epsilon\sqrt{\epsilon^2+\Delta^2}
+\Delta^2\ln\!\left(
\frac{\epsilon+\sqrt{\epsilon^2+\Delta^2}}{\Delta}
\right)
\right].
\end{align}
where only in the last step $\Delta\ll1$ was assumed. Keeping only the leading term in the endpoint expansion provides a uniform approximation that captures the logarithmic dependence on $\Delta$ as $\Delta\to 0$. The expression above coincides with the regularizing term appearing in Eq.~(\ref{Eq:ET}).

\section{Distortions in the transverse direction}
\label{App:distortions}

In the main text we restricted the lattice distortion to the longitudinal direction. To assess the validity of this approximation, we extend the model by introducing a transverse dimerization channel characterized by the order parameter $\Delta_\perp=\delta_\perp/2\gamma$. We find that transverse dimerization can become weakly favorable when the longitudinal Peierls instability is absent or very weak. However, once a sizeable longitudinal CDW develops, the energy gain associated with transverse dimerization is strongly suppressed. Consequently, in the anisotropic regime $\gamma\ll t$ considered throughout this work, the longitudinal distortion constitutes the dominant ordering channel.

\begin{figure*}
    \centering
    \includegraphics[width=1\linewidth]{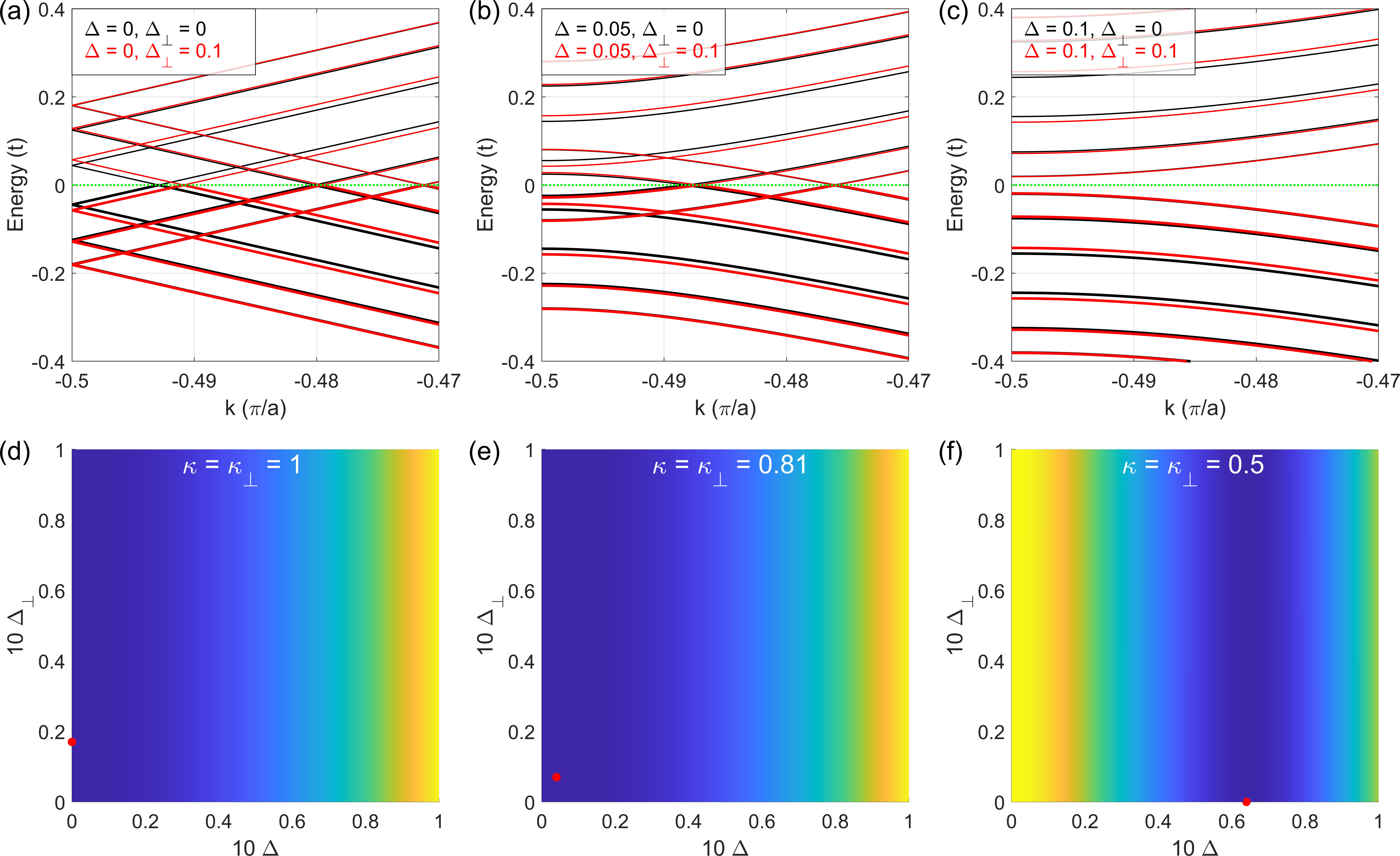}
    \caption{Band structure near the Fermi energy for increasing longitudinal dimerization $\Delta$, comparing the cases$\Delta_\perp=0$ (black) and $\Delta_\perp=0.1$ (red). The effect of transverse dimerization becomes progressively less relevant as the longitudinal Peierls gap develops. (d)-(f) Total energy in the $(\Delta,\Delta_\perp)$ plane for $\gamma=0.1t$ and $\kappa=\kappa_\perp=1, 0.81$ and $0.5$, respectively. Red dots highlights the position of the global energy minimum.}
    \label{fig:R1}
\end{figure*}

We assume an alternating interchain coupling of the form $\gamma_\pm\equiv\gamma\pm\delta_\perp/2$, which  enters the Hamiltonian as
\begin{align}
    \mathcal{H}= \mathbb{I}_N\otimes H_k+H_\perp\otimes\mathbb{I}_2+
    \Delta_\perp
    [
    S
    H_\perp]
    \otimes 
    \mathbb{I}_2,
    \nonumber
\end{align}
with $S=\mathrm{diag} \left[1,-1,1,\cdots,(-1)^{N-1}\right]$. We also include the transverse contribution to the elastic energy per unit cell,
\begin{align}
    E_{\mathrm{elastic}}&=
    NK x^2+
    (N-1)K_\perp z^2\\
    &=
    4N
    t^2\kappa\Delta^2+
    4(N-1)
    \gamma^2\kappa_\perp\Delta_\perp^2,\nonumber
\end{align}
where we introduce parameters analogous to the longitudinal counterpart, transverse rigidity $\kappa_\perp=K_\perp/\alpha_\perp^2$, with $\delta_\perp=-\alpha_\perp z$ and $z$ being the physical transverse displacement of the interatomic distance of the chains. We then compute the total energy by integrating all occupied states below the Fermi level, exactly as in the main text.

Figures \ref{fig:R1}(a)-(c) illustrate the evolution of the band structure with increasing longitudinal CDW order. In the weak-CDW regime, the transverse modulation lowers the electronic energy through small shifts of the occupied subbands. However, once a sizeable longitudinal Peierls gap develops, the additional spectral reconstruction induced by $\Delta_\perp$ produces an equal upward and downward shifts of the occupied states. As a consequence, the net electronic energy gain becomes negligible, leaving the elastic energy cost as the dominant contribution to the total energy.

To investigate how these three regimes are reflected in the phase diagram, we choose $\kappa=\kappa_\perp$ and $\gamma=0.1t$, and present the $(\Delta,\Delta_\perp)$ map of the total energy in panels (d), (e) and (f) of the same figure, for decreasing values of the rigidity. For high enough values of the rigidity, the phase with $\Delta=0$ is the most energetically favorable, as shown in the rightmost panel of Fig. \ref{fig:3} (a) for $\gamma=0.1t$ and $\kappa=1$. In this case, a small transverse distortion seems to be energetically preferred, as highlighted by the red dot in panel (d). However, the total energy of the distorted phase is about $\sim10^{-7}t$ smaller with respect to the phase without transverse distortion. For smaller values of the rigidity, panel (e) shows that the minimum energy corresponds to finite values of both $\Delta$ and $\Delta_\perp$. The panel (f) shows that for sufficiently small values of the rigidity, the strong intrachain dimerization lowers the total energy by approximately $10^{-2}$--$10^{-3}t$, which is four to five orders of magnitude larger than the energy gain associated with the transverse distortion. Thus, in this regime, the optimal value of $\Delta$ fully gaps the electronic structure, rendering the transverse dimerization energetically unfavorable since it no longer provides an additional electronic energy reduction. 

Finally, in the two-dimensional limit, CDW instabilities are commonly analyzed through the charge susceptibility \cite{gruner_density_1994, luckin_controlling_2024}. In this language, the alternating pattern $\gamma_+, \gamma_-, \gamma_+,\cdots$, corresponds to a transverse ordering vector $Q_y=\pi/a$. Our results indicate that, in the strongly anisotropic regime $\gamma\ll t$, the longitudinal Peierls gap reconstructs the electronic spectrum in such a way that the additional energy gain associated with a transverse gap becomes negligible. Consequently, the transverse distortion channel is strongly suppressed, which justifies the restriction to longitudinal distortions adopted in the main text.


\section{Phase diagram of 2N parallel-coupled chains}
\label{App:2N}
In the main text, we pointed out that all systems consisting of \(2N\) parallel-coupled chains share the same phase diagram. Here we rationalize this empirical observation by explicitly computing the critical inter-chain coupling \(\gamma_c\) above which the charge-density-wave (CDW) state becomes unstable. To this end, we evaluate the total-energy difference between the CDW and the normal state,
\begin{align}
\delta E_{\mathrm{T}}^{2N}(\Delta)\equiv
E_{\mathrm{T}}^{2N}(\Delta)-E_{\mathrm{T}}^{2N}(0).
\end{align}
For \(\Delta=0\) the total energy reads
\begin{align}
E_{\mathrm{T}}^{2N}(0)=
-\frac{8t}{\pi}N+\frac{4t}{\pi}\sum_{n=1}^N\epsilon_n^2,
\end{align}
where \(\epsilon_n\) accounts for the band-edge shift induced by the inter-chain hybridization. For finite \(\Delta\) we obtain
\begin{widetext}
\begin{align}
E_{\mathrm{T}}^{2N}(\Delta)=&
8Nt^2\kappa\Delta^2
-\frac{8t}{\pi}\sum_{n=1}^N
\int_{0}^{\pi/2-\epsilon_n}d\phi\,
\sqrt{1-\left(1-\Delta^2\right)\sin^2\phi}.
\end{align}
Expanding the elliptic integral for \(\Delta\ll 1\), this expression can be written as
\begin{align}
E_{\mathrm{T}}^{2N}(\Delta)=&
8Nt^2\kappa\Delta^2
-\frac{8t}{\pi}N
\left[
1+\frac{\Delta^2}{2}
\left(
\ln\frac{4}{\Delta}-\frac{1}{2}
\right)
\right]
+\frac{4t}{\pi}
\sum_{n=1}^N
\left[
\epsilon_n\sqrt{\epsilon_n^2+\Delta^2}
+\Delta^2
\ln\left(
\frac{\epsilon_n+\sqrt{\epsilon_n^2+\Delta^2}}{\Delta}
\right)
\right].
\end{align}
\end{widetext}

For odd $N$ parallel-coupled chains at half filling, one pair of bands has \(\epsilon=0\), and the logarithmic singularity survives, driving a Peierls-like instability (see light blue region in Fig.\ref{fig:3}). For an even number of coupled chains, however, the transverse dispersion shift
\begin{align}
\epsilon_n = 2\gamma\cos\left(\frac{n\pi}{2N+1}\right),
\end{align}
is never zero for integer \(n\in[1,N]\), and the divergence is therefore regularized. Since the CDW gap is the smallest energy scale in our problem (\(\Delta\sim 10^{-6}t\)), we expand the total energy to leading order in \(\Delta\). After straightforward algebra we obtain
\begin{align}\label{Eq:deltaE}
\delta E_{\mathrm{T}}^{2N}(\Delta)=
\frac{4t\Delta^2}{\pi}\left[
2Nt\kappa\pi+
N\left(1-\ln 2\right)
+
\sum_{n=1}^N\ln\left(\frac{\epsilon_n}{t}\right)
\right].
\end{align}
Using \(\epsilon_n\simeq (\gamma/t)\cos\!\left(\frac{n\pi}{2N+1}\right)\), the transition line is obtained by imposing \(\delta E_{\mathrm{T}}^{2N}(\Delta)=0\), which yields
\begin{widetext}
\begin{align}\label{Eq:condition}
2t\kappa\pi+
\left(1-\ln 2\right)
+\ln\left(\frac{\gamma}{t}\right)
+\frac{1}{N}
\sum_{n=1}^N
\ln\left[
\cos\left(\frac{n\pi}{2N+1}\right)
\right]
=0.
\end{align}
\end{widetext}

At first sight, Eq.~(\ref{Eq:condition}) appears to depend on \(N\) through the last term. However, this dependence cancels exactly. To show this, we write the polynomial \(z^{2N+1}+1\) as a product over its roots,
\begin{align}
z^{2N+1}+1=
\prod_{j=0}^{2N}\left(z-e^{i\theta_j}\right),
\qquad
\theta_j=\frac{(2j+1)\pi}{2N+1}.
\end{align}
Evaluating at \(z=1\) gives
\begin{align}
2=\prod_{j=0}^{2N}(1-e^{i\theta_j}).
\end{align}
The factor with \(j=N\) corresponds to \(\theta_N=\pi\) and thus equals \(1-e^{i\pi}=2\). Factoring it out, we obtain
\begin{align}
1=\prod_{j\neq N}(1-e^{i\theta_j}).
\end{align}
Using \(\theta_{2N-j}=2\pi-\theta_j\), we can pair complex conjugates and write
\begin{align}
1&=\prod_{j=0}^{N-1}(1-e^{i\theta_j})(1-e^{-i\theta_j})
=\prod_{j=0}^{N-1}4\sin^2\left(\frac{\theta_j}{2}\right).
\end{align}
Taking the square root yields
\begin{align}
\prod_{j=0}^{N-1}\sin\left(\frac{\theta_j}{2}\right)=2^{-N}.
\end{align}
Finally, since
\begin{align}
\sin\left(\frac{\theta_j}{2}\right)
=
\sin\left[\frac{(2j+1)\pi}{2(2N+1)}\right]
=
\cos\left[\frac{(j+1)\pi}{2N+1}\right],
\end{align}
we obtain the identity
\begin{align}
\prod_{n=1}^{N}
\cos\left(\frac{n\pi}{2N+1}\right)=2^{-N}.
\end{align}
Therefore,
\begin{align}
\frac{1}{N}\sum_{n=1}^N
\ln\left[
\cos\left(\frac{n\pi}{2N+1}\right)
\right]
=
\frac{1}{N}\ln\left(2^{-N}\right)
=-\ln 2,
\end{align}
which is independent of \(N\). This proves that the critical coupling \(\gamma_c\) obtained from Eq.~(\ref{Eq:condition}) is the same for all \(2N\)-chain systems, consistent with the \(N\)-independent phase diagram reported in the main text.

\section{Zig-zag and nematic phases at half-filling}
\label{App:zzg_vs_nem}

In this appendix we provide a simple perturbative argument showing that, at half filling, the zig-zag configuration is energetically favored over the nematic one. We focus on the minimal case of $N=3$ skew-coupled chains and evaluate the Hamiltonian at the Brillouin-zone corner $k=-\pi/2a$, where the Peierls distortion is expected to open a gap.

At $k=-\pi/2a$, the nematic and zig-zag Hamiltonians read
\begin{subequations}
\begin{align}
    H^{\mathrm{nem}}_{k=\frac{-\pi}{2a}}=&\left(
    \begin{matrix}
        \delta\sigma_x&T&0\\
        T^{\dagger}&\delta\sigma_x&T\\
        0&T^{\dagger}&\delta\sigma_x
    \end{matrix}
    \right),\\
    H^{\mathrm{zzg}}_{k=\frac{-\pi}{2a}}=&
    \left(
    \begin{matrix}
        \delta\sigma_x&T&0\\
        T^{\dagger}&\delta\sigma_x&T^{\dagger}\\
        0&T&\delta\sigma_x
    \end{matrix}
    \right),
\end{align}
\end{subequations}
where the inter-chain coupling is
\begin{align}
    T=\gamma
    \left(
    \begin{matrix}
        1&-1\\1&1
    \end{matrix}
    \right)=\gamma(\sigma_x-i\sigma_y).
\end{align}
We consider the Peierls term as a small perturbation,
\begin{align}
    V=\delta\,\mathrm{diag}(\sigma_x,\sigma_x,\sigma_x),
\end{align}
and define the unperturbed Hamiltonians ($\delta=0$),
\begin{subequations}
\begin{align}
    H^{\mathrm{nem}\,(0)}=&\left(
    \begin{matrix}
        0&T&0\\
        T^{\dagger}&0& T\\
        0& T^{\dagger}&0
    \end{matrix}
    \right),\\
    H^{\mathrm{zzg}\,(0)}=&    
    \left(
    \begin{matrix}
        0&T&0\\
         T^{\dagger}&0& T^{\dagger}\\
        0& T&0
    \end{matrix}
    \right).
\end{align}
\end{subequations}
Both $H_{\mathrm{nem}}^{(0)}$ and $H_{\mathrm{zzg}}^{(0)}$ host a twofold degenerate eigenvalue at $E^{(0)}=0$. The corresponding eigenstates can be constructed explicitly by taking a spinor $\phi$ on the first chain and requiring the middle-chain component to vanish. Then the zero-energy eigenvectors read,
\begin{subequations}
\begin{align}
    \Psi_{\mathrm{nem}}=
    \frac{1}{\sqrt{2}}
    \left(
    \begin{matrix}
        \phi\\
        0\\
        -i\sigma_y\phi
    \end{matrix}
    \right),\quad
    \Psi_{\mathrm{zzg}}=
    \frac{1}{\sqrt{2}}
    \left(
    \begin{matrix}
        \phi\\
        0\\
        -\phi
    \end{matrix}
    \right),\nonumber
\end{align}
\end{subequations}
Since the zero-energy eigenspace is two-dimensional, the first-order spectrum is obtained by diagonalizing the perturbation restricted to this subspace \cite{sakurai_modern_2017}. For the {\it{nematic configuration}}, the matrix elements of the first-order perturbation in the degenerate subspace are
\begin{align}
    V_{\mathrm{eff}}^{(\mathrm{nem})}
    =&
    \langle\Psi_{\mathrm{nem}}\vert
    V
    \vert\Psi_{\mathrm{nem}}\rangle\\
    =&
    \frac{\delta}{2}
    \left[\sigma_x +    
    (-i\sigma_y)^\dagger\sigma_x(-i\sigma_y)
    \right]
    =0.
\end{align}
Therefore the degeneracy at $E^{(0)}=0$ is not lifted at first order in $\delta$ in the nematic configuration, $E_{\mathrm{nem}}^{(1)}=0$. For the {\it{zig-zag configuration}}, however, the effective perturbation becomes
\begin{align}
    V_{\mathrm{eff}}^{(\mathrm{nem})}
    =&
    \langle\Psi_{\mathrm{nem}}\vert
    V
    \vert\Psi_{\mathrm{nem}}\rangle\\
    =&
    \frac{\delta}{2}
    \left[\sigma_x +    
    \sigma_x
    \right]
    =\delta \sigma_x.
\end{align}
Diagonalizing the matrix above yields the first-order energies $E_{\mathrm{zzg}}^{(1)} = \pm \delta$. Hence, in the zig-zag configuration the twofold degeneracy is lifted already at first order, producing a linear-in-$\delta$ splitting at the Brillouin-zone corner. A larger splitting produces a larger lowering of the occupied energies, which always translates into a stronger Peierls CDW phase. Since the nematic configuration shows no first-order splitting, while the zig-zag configuration splits as $2\delta$ at $k=-\pi/2a$, we conclude that the zig-zag phase is energetically preferred at (and near) half filling.

\newpage

\bibliographystyle{unsrt}

\bibliography{references}

\end{document}